%% file: manuscript.tex
\documentclass[twocolumn]{aastex62}
\usepackage{amsmath}
\usepackage{comment}
\usepackage{xcolor}

\input{shortcuts.tex}

\shorttitle{AN INDEPENDENT ANALYSIS OF THE SIX RECENTLY CLAIMED EXOMOON CANDIDATES}
\shortauthors{Kipping}
\begin{document}

\title{AN INDEPENDENT ANALYSIS OF THE SIX RECENTLY CLAIMED EXOMOON CANDIDATES}

\correspondingauthor{David Kipping}
\email{dkipping@astro.columbia.edu}

\author[0000-0002-4365-7366]{David Kipping}
\affil{Department of Astronomy,
Columbia University,
550 W 120th Street,
New York, NY 10027, USA}
\affil{Center for Computational Astophysics,
Flatiron Institute,
162 5th Av.,
New York, NY 10010, USA}



\begin{abstract}
It has been recently claimed that KOIs-268.01, 303.01, 1888.01, 1925.01,
2728.01 \& 3320.01 are exomoon candidates, based on an analysis of their
transit timing. Here, we perform an independent investigation, which is framed
in terms of three questions: 1) Are there significant excess TTVs? 2) Is there a
significant periodic TTV? 3) Is there evidence for a non-zero moon mass? We
applied rigorous statistical methods to these questions alongside a re-analysis
of the Kepler photometry and find that none of the KOIs satisfy these three
tests. Specifically, KOIs-268.01 \& 3220.01 pass none of the tests and
KOIs-303.01, 1888.01 \& 1925.01 pass a single test each. Only KOI-2728.01
satisfies two, but fails the cross-validation test for predictions. Further,
detailed photodynamical modeling reveals that KOI-2728.01 favours a negative
radius moon (as does KOI-268.01). We also note that we find a significant
photoeccentric for KOI-1925.01 indicating an eccentric orbit of
$e>(0.62\pm0.06)$. For comparison, we applied the same tests to Kepler-1625b,
which reveals that 1) and 3) are passed, but 2) cannot be checked with the
cross-validation method used here, due to the limited number of available
epochs. In conclusion, we find no compelling evidence for exomoons amongst the
six KOIs. Despite this, we’re able to derive exomoon mass upper limits versus
semi-major axis, with KOI-3220.01 leading to particularly impressive
constraints of $M_S/M_P < 0.4$\% [2\,$\sigma$] at a similar relative semi-major to that
of the Earth-Moon.
\end{abstract}

\keywords{planets and satellites: detection --- methods: statistical}

\section{Introduction}

\input{intro.tex}

\section{Target Data}
\label{sec:data}

\input{data.tex}

\section{Analysis}
\label{sec:analysis}

\input{fits.tex}

\section{Discussion}
\label{sec:discussion}

\input{discussion.tex}

\section*{Acknowledgements}

DMK is supported by the Alfred P. Sloan Foundation. Thanks to Alex
Teachey and the Cool Worlds team for useful discussions, and to the
anonymous reviewer for a constructive report. Special thanks to Tom Widdowson,
Mark Sloan, Laura Sanborn, Douglas Daughaday, Andrew
Jones, Jason Allen, Marc Lijoi, Elena West \& Tristan Zajonc.

%\newpage



\appendix

\section*{}

\begin{figure*}[h!]
\begin{center}
\includegraphics[width=17.0cm,angle=0,clip=true]{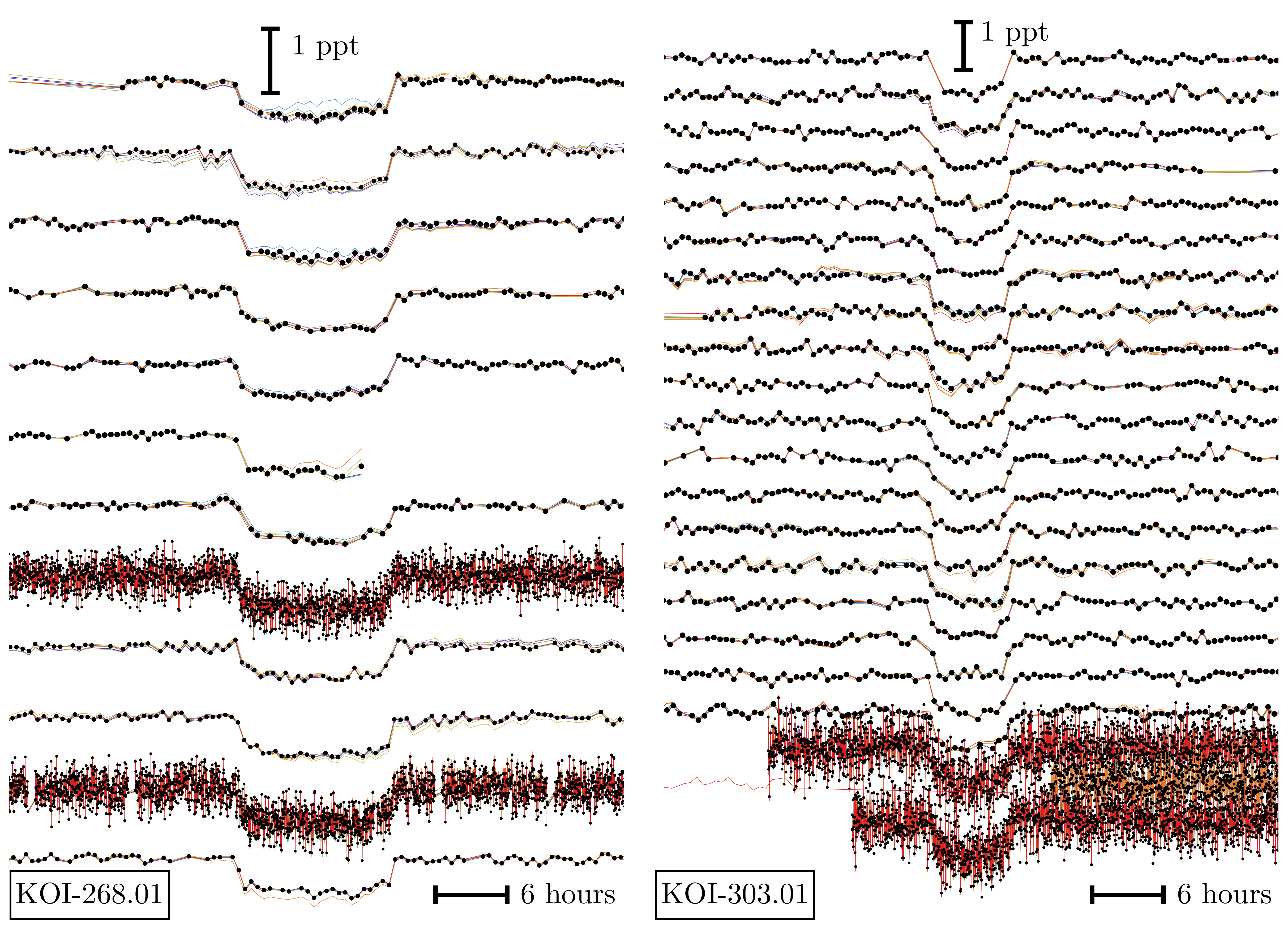}
\caption{
Method marginalized detrended light curves for KOI-268.01 and KOI-303.01.
The eight colored continuous lines show the eight different independent
detrendings of each epoch, which are then combined together to form the
method marginalized time series (black points).
}
\label{fig:lc1}
\end{center}
\end{figure*}

\begin{figure*}
\begin{center}
\includegraphics[width=17.0cm,angle=0,clip=true]{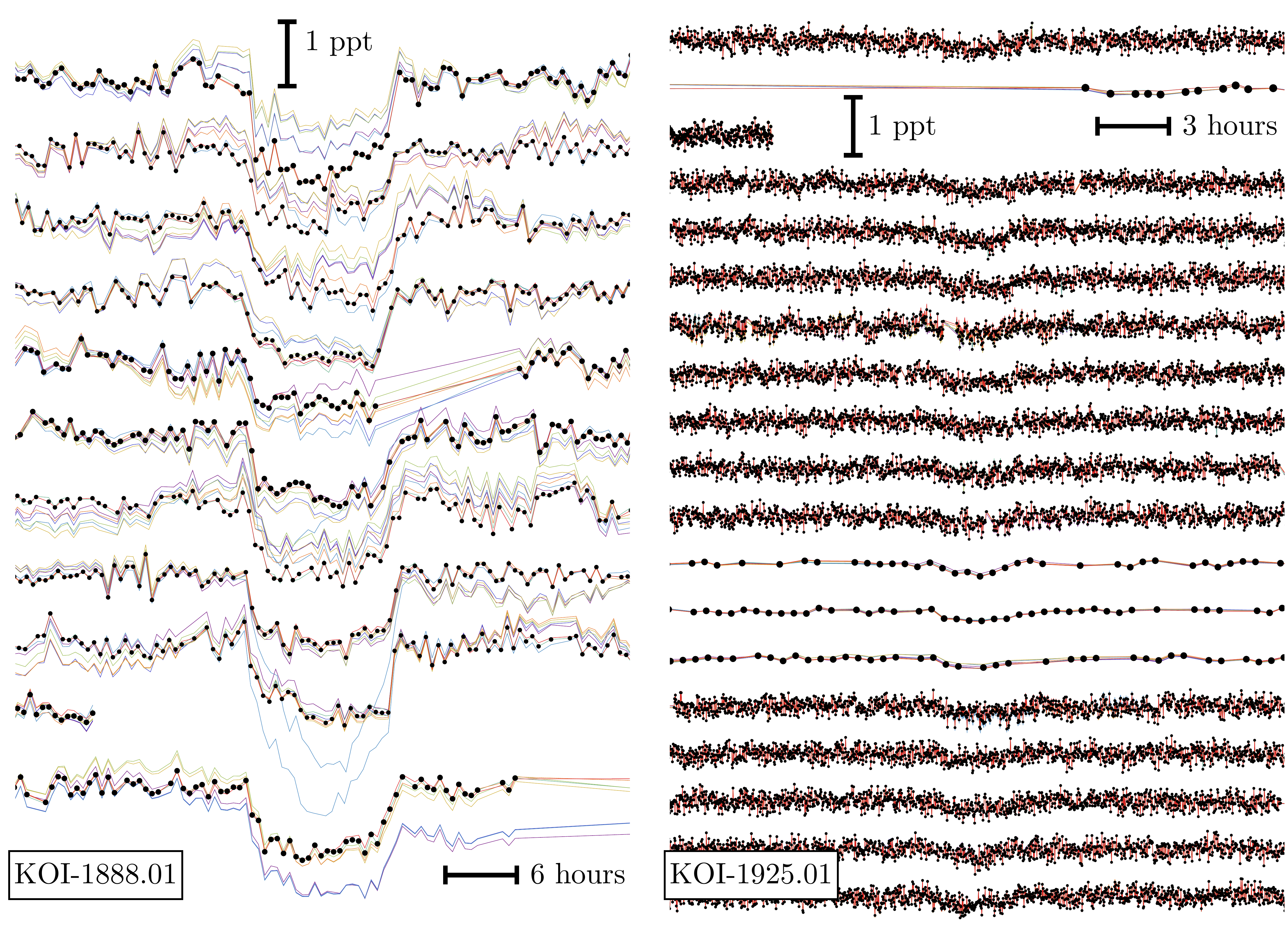}
\caption{
Method marginalized detrended light curves for KOI-1888.01 and KOI-1925.01.
The eight colored continuous lines show the eight different independent
detrendings of each epoch, which are then combined together to form the
method marginalized time series (black points).
}
\label{fig:lc2}
\end{center}
\end{figure*}

\begin{figure*}
\begin{center}
\includegraphics[width=17.0cm,angle=0,clip=true]{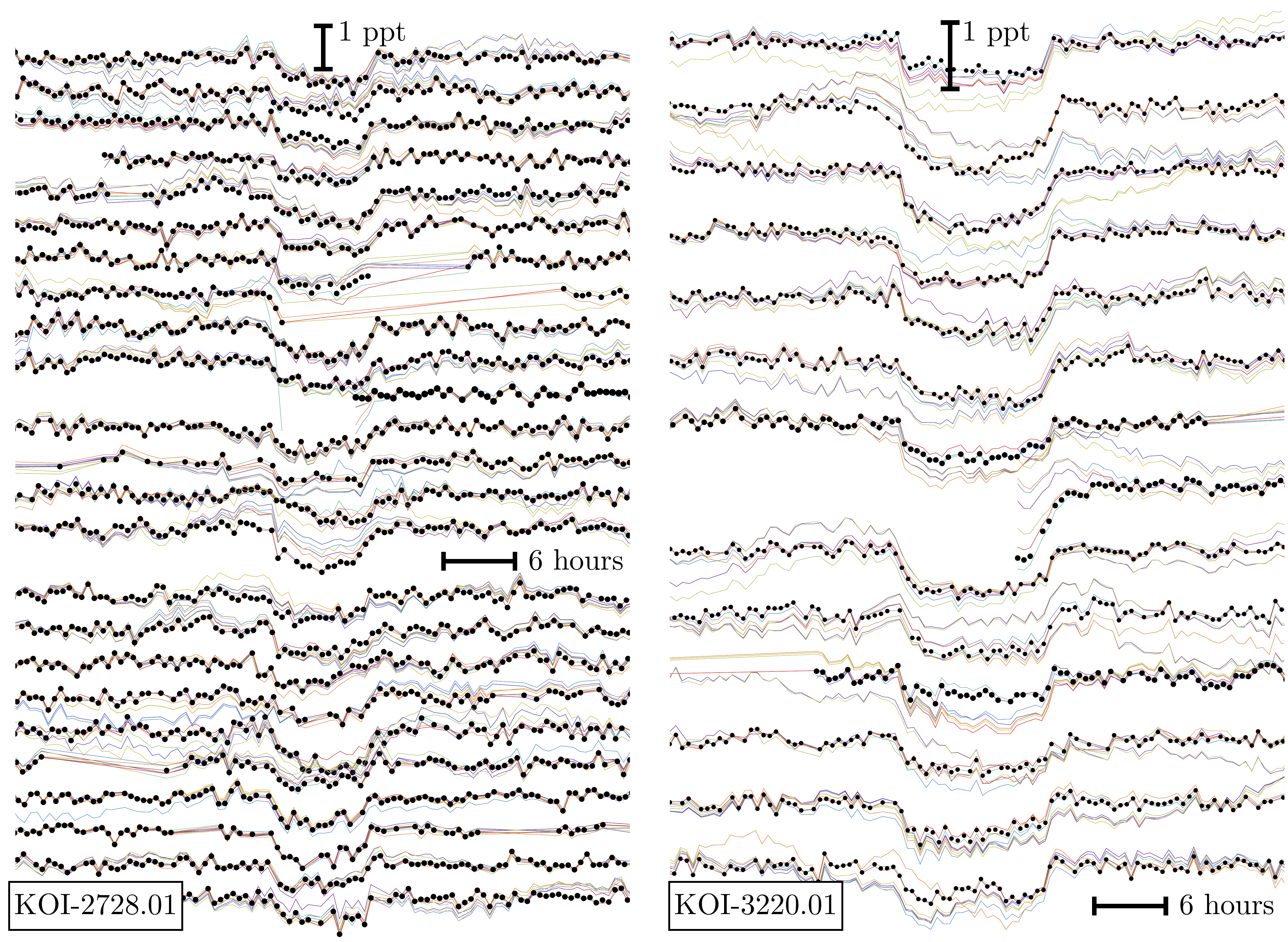}
\caption{
Method marginalized detrended light curves for KOI-2728.01 and KOI-3220.01.
The eight colored continuous lines show the eight different independent
detrendings of each epoch, which are then combined together to form the
method marginalized time series (black points).
}
\label{fig:lc3}
\end{center}
\end{figure*}

\input{apptables.tex}

\end{document}

%% file: shortcuts.tex
\newcommand{\kepler}{{\it Kepler}}

\newcommand{\multi}{{\tt MultiNest}}
\newcommand{\luna}{{\tt LUNA}}

\newcommand{\isochrones}{{\tt isochrones}}
\newcommand{\wwwcoolworlds}{\href{https://github.com/CoolWorlds/ttvposteriors}{this URL}}

%% file: intro.tex
It has been recently proposed that six Kepler Objects Interest (KOI) host
candidate exomoons in \citet{fox:2020}. Given the paucity of these objects in
the literature, this would represent a major increase in the number of
known candidates, as of the time of writing. For this reason, we here
provide an independent analysis of the moon hypothesis for these
six: KOIs-268.01, 303.01, 1888.01, 1925.01, 2728.01 \& 3320.01.

It has been proposed that exomoons could be discovered through a myriad of
approaches, such as pulsar timing \citep{lewis:2008}, microlensing
\citep{han:2002} and spectroscopy \citep{williams:2004}, but the transit
method is somewhat unique in offering the ability to measure the mass
and radius of potential moons (see review by \citealt{heller:2014}). The
mass is available by the study of transit timing effects imparted by the
moon upon the planet, which include transit timing variations (TTVs;
\citealt{sartoretti:1999}), velocity induced transit duration variations
(TDV-Vs; \citealt{kipping:2009a}), transit impact parameter induced
transit duration variations (TDV-TIPs; \citealt{kipping:2009b}), and
ingress/egress asymmetries \citep{luna:2011}. Whilst all of these are
generally present, TTVs typically offer the most detectable signal and
are the more commonly cataloged timing effect (e.g. see
\citealt{mazeh:2013}).

The case for TTVs is strengthened when one considers that they appear common
amongst KOIs \citep{holczer:2016}, potentially indicating a large number of
unrevealed exomoons. Indeed, it was recently shown that of 2416 KOIs
with a model preference for a periodic TTV, 2198 of them exhibit TTVs and TDVs consistent with an exomoon
\citep{impossible}. In that paper, amongst the 2198 aforementioned cases, one
finds that KOIs-268.01, 303.01, 1888.01, 1925.01, 2728.01 and 3320.01 are indeed all
listed in their Table~1 as being fully consistent with an exomoon. Although,
the authors refrained from describing these as exomoon candidates, nor
indeed any of the other 2192 cases.

A basic reason for this is that although the TDVs were consistent with an
exomoon, no significant detection of them had been made; only often very
tentative evidence for TTVs existed. Certainly, in many fields a single
type of observational information can be sufficient to securely claim a
detection, but the unique challenge facing TTVs is that a considerable
number of non-exomoon phenomena can equally cause TTVs. These include,
but are not limited to,
exotrojans \citep{ford:2007},
parallax effects \citep{scharf:2007},
eccentricity variations \citep{kipping:2008},
apsidal precession \citep{jordan:2008},
star spots \citep{alonso:2008},
stellar proper motion \citep{rafikov:2009},
planetary in-fall \citep{hellier:2009a},
the Applegate effect \citep{applegate:1992,watson:2010},
stellar binarity \citep{montalto:2010},
cadence stroboscoping \citep{szabo:2013},
horseshoe companions \citep{vokrouhlicky:2014},
planet-planet conjunctions \citep{nesvorny:2014},
and, near mean motion resonant planets \citep{agol:2005,holman:2005}.
Planet-planet interactions are particularly common, given the
abundance of packed planetary systems found amongst the \kepler\
sample \citep{winn:2015}. On this basis, the existence of TTVs, even
periodic TTVs, can be perilous ground upon which to solely base an exomoon
claim.

The six candidates claimed by \citet{fox:2020} seem to exhibit TTVs, and
thus are indeed consistent with one observational effect of exomoons then.
Further, the authors selected targets for exomoons that are plausibly
dynamically stable, and honed in on the highest signal-to-noise transits
available. That latter point is particularly important since one might
expect these KOIs to enable particularly sensitive searches for exomoons.
A search for exomoons is thus interesting around these KOIs in its
own right.

Accordingly, in this work, we will interrogate the claim of
\citet{fox:2020} for each of the six KOIs argued to be exomoon candidates.

%% file: data.tex
\subsection{Background}
\label{sub:datamotivation}

In \citet{fox:2020}, the authors relied on a catalog of transit timing
measurements presented in \citet{holczer:2016}. As the \kepler\ light curves
upon which these transit times are derived are publicly available, and the
number of objects is fairly small, we elected to derive
our own own transit timing estimates.

There are several reasons for doing this. First, the \citet{holczer:2016}
transit times are the product of an automated analysis, which made several
approximations to expedite their calculation. For example, uncertainties
are assigned using an empirical relation rather than actually being formally
determined for each object \citep{holczer:2016}. Second, the analysis was
conducted prior to the final \kepler\ Data Release, DR25, and thus does not
use the most up to date reduction of the \kepler\ light curves (nor indeed
short cadence data where available). Third, the magnitude of the the claim of
\citet{fox:2020} warrants a careful independent analysis to interrogate their
hypothesis.

\subsection{Method marginalized light curve detrending}
\label{sub:methmarg}

To detrend the \kepler\ light curves, we follow the approach of
\citet{teachey:2018} and detrend the light curve multiple ways through
method marginalized detrending. Of the five detrending approaches used
in \citet{teachey:2018}, we use the same set here except we drop the
median filtering approach, as it was found to be the least reliable
in that work (see their Figure~S7).

We obtained the Simple Aperture Photometry (SAP) and Pre-Data search
Conditioning (PDC) DR25 %\footnote{Processed by v9.3 of the SOC pipeline.}
photometric time series from MAST for each KOI.
Short-cadence (SC) data is used with preference over long-cadence (LC), whenever
available. We first applied a removal of outliers based on any error flags in
the {\tt fits} file, or 4\,$\sigma$ deviations from a 21-point rolling median.
Each transit epoch is then detrended independently (where an epoch is centered
on the time of transit minimum and spans $\pm0.5P_P$) with all four algorithms,
on both the SAP and PDC time series, giving a total of eight light curves per
epoch per target (see Appendix for light curves).

We next generated 1000 fake light curves for each method and epoch assuming
pure Gaussian noise, in order to test how whitened each light curve is. First,
we computed a simple Durbin-Watson statistic \citep{durbin:1950} and rejected
any light curves which exhibit autocorrelation more than 2\,$\sigma$ deviant
from white, as characterized by the fake light curve population. Second, we
binned the light curves into progressively larger bins, computed an RMS, and
then fitted a gradient through the log-log plot of bin size versus RMS. This
was done for every fake light curve as well as the real, allowing us to again
reject any light curves for which the binning properties are more than
2\,$\sigma$ deviant from white noise behaviour.

The $\leq 8$ surviving light curves (per epoch) were then combined
into a so-called method marginalized light curve by calculating an inter-method
median at each time stamp, and propagating the standard deviation between
methods into that data point's error budget through quadrature summation. Lastly,
each epoch is appended to a single file that is used for the subsequent
light curve fits described in Section~\ref{sec:analysis}.

%% file: fits.tex
The claim of \citet{fox:2020} is that KOIs-268.01,303.01, 1888.01,
1925.01, 2728.01 and 3320.01 are ``exomoon candidates'', which
is based upon an analysis of the transit times published by
\citet{holczer:2016}. Exomoons of transiting planets will also
transit their parent star, presenting an additional piece of information
that may be used to infer their presence (e.g. \citealt{HEK2}). Nevertheless,
we focus on transit timing in what follows since that is the basis upon
which the claim of \citet{fox:2020} was made.

To this end, we consider three basic questions for each of the six KOIs
under consideration:

\begin{itemize}
\item[{\textbf{Q1]}}] Are there statistically significant TTVs?
\item[{\textbf{Q2]}}] Is there a statistically significant periodic TTV?
\item[{\textbf{Q3]}}] Do the observations support a statistically significant non-zero moon mass?
\end{itemize}

In the following three subsections, we tackle each of these questions
in-turn, and then apply the same tests to a previously announced exomoon candidate,
Kepler-1625b in the final part of this section.

\subsection{Q1 - Are there significant TTVs?}
\label{sub:Q1}

\subsubsection{Inferring the transit times}

To derive TTVs, we first modeled the transit light curve using the
\citet{mandel:2002} formalism with quadratic limb darkening (using the
$q_1$-$q_2$ parameterization of \citealt{q1q2}), and the light curve
integration scheme of \citet{binning} to account for LC smearing. Two
versions of this model were considered against the data. Model $\mathcal{P}$
assumes a linear ephemeris characterized by an orbital period $P$ and
reference time of transit minimum, $\tau_0$. Model $\mathcal{T}$ allows
for TTVs by giving each epoch a unique time of transit minimum, $\tau_i$.

The models are regressed to the method marginalized light curves using
a multimodal nested sampling algorithm, \multi\ \citep{feroz:2008,feroz:2009},
providing marginal likelihoods and posterior samples.
Priors were set to be uniform for any ephemeris parameters, to within
$\pm 0.5$\,days of the NASA Exoplanet Archive (NEA) ephemeris
\citep{akeson:2013}. The stellar density ($\rho_{\star}$) used a log-uniform
prior from $10^{-3}$\,g\,cm$^{-3}$ to $10^{+3}$\,g\,cm$^{-3}$, impact
parameter ($b$) was uniform from 0 to 2, the ratio-of-radii ($p$) was uniform
from 0 to 1, as were the limb darkening coefficients $q_1$ \& $q_2$. Formally,
the orbit is circular but the ability for the stellar density to vary
effectively allows for eccentric orbits since this allows the velocity
of the planet to vary. The photometry was modeled with a
normal likelihood function, which is justified on the basis that our
detrending pre-whitened the data with explicit tests for gaussianity (see
Section~\ref{sub:methmarg}).

Since model $\mathcal{T}$ assigns a unique $\tau_i$ to each epoch,
this can lead to short-period planets having a large number of total
free parameters to explore, which impedes parameter exploration. To
circumvent this, we segmented such fits into two subsets of ${\sim}10$
epochs, which was necessary for KOIs-303.01, 1925.01 \& 2728.01.

\subsubsection{Comparison to times used by \citet{fox:2020}}

\begin{figure*}
\begin{center}
\includegraphics[width=17.0cm,angle=0,clip=true]{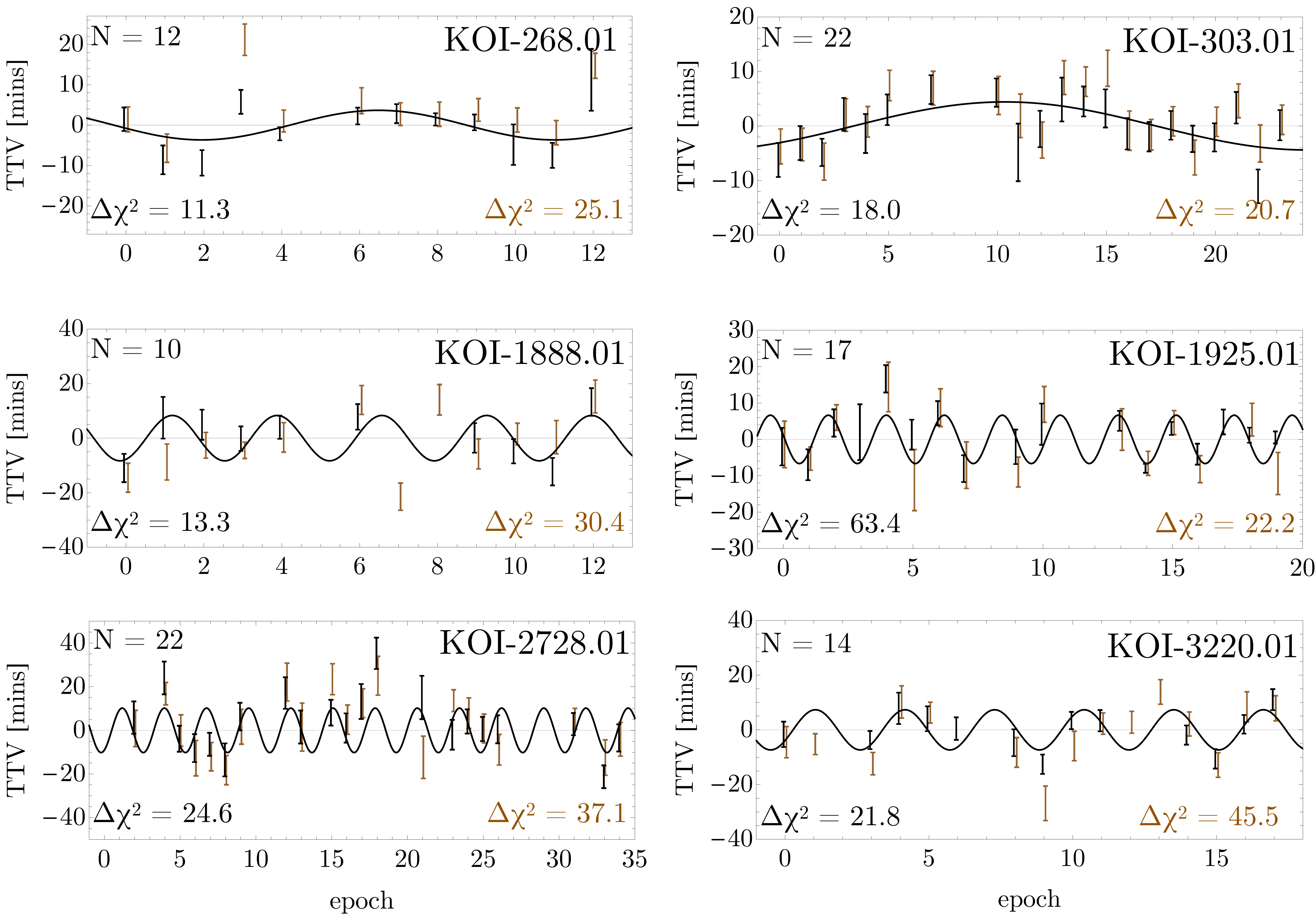}
\caption{
TTVs for the six claimed exomoon candidate hosts of \citet{fox:2020}, with our
own measurements in black and those of \citet{holczer:2016} in brown. We
overplot the best-fitting sinusoid with the $\chi^2$ improvement shown in the
lower-left corner (and similarly for that of \citet{holczer:2016} in the
lower-right, although we do not plot the associated sinusoid).
}
\label{fig:TTVs}
\end{center}
\end{figure*}

From this process, we obtained marginalized
posterior distributions for the $\tau_i$ parameters for each KOI,
which are summarized in the Appendix (Tables~\ref{tab:koi268}-\ref{tab:koi3220})
and made available at \wwwcoolworlds. We derived summary
statistics for each epoch by computing the median and $\pm$38.1\% range. Transit
times were converted to TTVs by subtracting the maximum a-posteriori ephemeris
derived from model $\mathcal{P}$. For all KOIs, the bulk of the TTVs points
exhibited deviations no larger than approximately half an hour, and thus we
excluded any points greater than an hour as outliers. These TTVs are shown in
Figure~\ref{fig:TTVs}, alongside those of \citet{holczer:2016} (and used
by \citealt{fox:2020}) for comparison.

There are noticeable differences between our TTVs
and those of \citet{holczer:2016}. For every KOI except KOI-1925.01, we
find that the $\Delta \chi^2$ improvement of a best sinusoidal fit versus
a linear ephemeris is decreased when using our TTVs (see values inset
in panels of Figure~\ref{fig:TTVs}) - thus largely attenuating the
significance of any TTVs.

\subsubsection{The challenge of defining TTV significance}

Equipped with our new transit times, let us ask whether there
are statistically significant variations - a subtle and non-trivial task.
One might consider a metric such as the reduced chi-squared, as
utilized by \citet{fox:2020}, but since the model is non-linear then that
metric is inappropriate \citep{andrae:2010}.

One might consider comparing the marginal likelihoods evaluated
from \multi\ for models $\mathcal{T}$ and $\mathcal{P}$. However, as noted
earlier, model $\mathcal{T}$ over-parameterizes the
problem here\footnote{Whilst this over-parameterization is indeed an issue for
marginal likelihoods, it's highly useful for posterior inference,
since the approach is agnostic as to the cause/shape of possible TTVs and
thus lacks any strong model conditionality.} leading to overly conservative
estimates for model $\mathcal{T}$. Indeed,
for all six KOIs, model $\mathcal{P}$ would be favoured using this approach.
% Removed to make more concise at the request of the reviewer...
%Another criticism of this approach is that the marginal likelihoods are
%sensitive to the priors, and here our priors are non-objectively selected
%- e.g. we somewhat arbitrarily used $\pm 0.5$\,days uniform priors for the
%transit times. We briefly note that this is less of a concern for the later
%photodynamic moon fits where the parameters have well-defined physical
%boundaries e.g. moon mass ratio is uniform from $0 \to 1$.

Instead then, one might consider evaluating some statistical measures on the
derived TTVs, such as the Bayesian Information
Criterion (BIC;\citealt{schwarz:1978}). However, those numbers
are summary statistics derived from a posterior, and thus applying statistical
tests to them is a) lossy, and b) demands certain
approximate assumptions. It is \textit{lossy} because when one adopts summary
statistics of a marginalized distribution, one ignores the full, rich detail of
the joint posterior shapes. To avoid such losses, it is preferable to make
inferences on the rawest data product which is practical (e.g. see \citet{hogg:2010}
for an analogous problem with eccentricities), which in our case would be the
photometric light curves. Applying
a test like the BIC to summary statistics is also \textit{approximate}, because it
requires an estimate of the maximum likelihood of a hypothesized model, and if that
likelihood is derived from summary statistics, then some approximation about the
likelihood function describing those summary statistics is necessary (e.g.
independent Gaussians).

\subsubsection{TTV significance tests with the photometry directly}

A better solution, then, is to apply model comparison tests on
the light curve products, but to avoid using the marginal likelihood
due to the parameterization problem of model $\mathcal{T}$ (ultimately that
issue is resolved with the photodynamics analysis in the next subsection).
Another important limitation is that $\mathcal{T}$ has no predictive power for
a held-out epoch, and thus we cannot directly use cross-validation either, at
least for Q1.

A basic quantity we can rely on is the maximum likelihood of the light curve
fits, $\hat{\mathcal{L}}$. Since $\mathcal{P}$ is, by definition, a nested
model of the more complex model $\mathcal{T}$, then $\hat{\mathcal{L}}$ will
always be greater for $\mathcal{T}$ - the real question is whether the
improvement outweighs the expense of the additional
complexity that model entails. A common tool for assessing this is the BIC,
given by $k \log n - 2\log\hat{\mathcal{L}}$,
where $k$ is the number of parameters estimated by the model and $n$ is the
number of data points\footnote{And also note that log is natural, unless
stated otherwise.}. %It should be noted that BIC is only appropriate when
%$n \gg k$ \citep{giraud:2015}, which is certainly true when applied to the
%light curve but again would be questionable if applied to the derived transit
%times.
Whilst $\hat{\mathcal{L}}$ and $n$ are well-defined, we again run into
an obstacle with $k$. If we set $k$ as the genuine number of free parameters
in the model (i.e. with each epoch requiring an additional parameter), it will
again over-parametrize the problem. The hypothesis of \citet{fox:2020} is an
exomoon, whose influence on the TTVs can be fully parameterized by just six
additional parameters \citep{luna:2011}, assuming a circular orbit moon as
expected due to rapid tidal circularization \citep{porter:2011}. All six
KOIs under consideration here include more than six transit epochs and thus
employ more parameters than is necessary to explain an exomoon (or indeed
a perturbing planet). On that basis,
we would expect a much more optimistic case for TTVs by using
$k_{\mathcal{T}}=6+k_{\mathcal{P}}$, and indeed a solution well-motivated by
our understanding of orbital parameterization. Accordingly, we argue that tests
for TTV signficance, when a single orbital component is hypothesized, should
use $k_{\mathcal{T}}+k_{\mathcal{P}} \to
\mathrm{min}[k_{\mathcal{T}}+k_{\mathcal{P}},6+k_{\mathcal{P}}]$, where the
minimum function accounts that in some cases we have less than 6 epochs.% and
%thus imposing 6 additional free parameterizations would be an
%over-parameterization.

% Removed the Bayesian Complexity calculation to make the paper more concise,
% as requested by the reviewer
%We briefly note that we also tried computing the Bayesian complexity, $\kappa$,
%of each model following the method described in \citet{kunz:2006} as a
%replacement for $k$, but found in every case that
%$\kappa_{\mathcal{T}}-\kappa_{\mathcal{P}}>6$, which would thus be a more
%conservative choice and decrease the case for TTVs\footnote{Specifically,
%we found $\kappa_{\mathcal{T}}-\kappa_{\mathcal{P}} = $6.5, 11.2, 6.9, 13.3,
%14.6, 8.2 for KOIs-268.01, 303.01, 1888.01, 1925.01, 2728.01 \& 3220.01
%respecitvely.} 

Proceeding as described, we find that that $\mathrm{BIC}_{\mathcal{P}} <
\mathrm{BIC}_{\mathcal{T}}$ for KOIs-268.01, 303.01, 1888.01 \& 3220.01,
indicating no TTVs, whereas KOI-1925.01 \& KOI-2728.01 do (see
Table~\ref{tab:models}).

\input{modeltable.tex}

We note a peculiarity about the two positive cases, though. Both are examples
of where it was necessary to segment the epochs into two groups, and so we are
also able to apply our statistical tests to each segment independently. In
doing so, we find that - for both KOIs - one segment shows positive evidence
but the other does not. For example, for KOI-1925.01, we obtain
$\mathrm{BIC}_{\mathcal{P}-\mathcal{T}} = -36.7$ for the first segment, but
$+75.7$ for the second. The first segment includes 7 short-cadence epochs out
of 10, whereas the second is 8 out of 9. Thus, this doesn't appear to offer a
good explanation for the large difference. Similarly, for KOI-2728.01 (which
has only long-cadence data), the first segment gives
$\mathrm{BIC}_{\mathcal{P}-\mathcal{T}} = 30.6$ but the second gives $-3.6$.
Since exomoons are expected to be strictly periodic signals, it is peculiar
for the significance to change versus time, implying a time-dependent
amplitude.

\subsection{Q2 - Is there a significant periodic TTV?}
\label{sub:Q2}

Having discussed whether the is statistical evidence for TTVs, we now
ask whether there is \textit{periodic} TTV embedded, as
expected for exomoons \citep{sartoretti:1999}.

We first note that model $\mathcal{T}$ has no predictive capacity for a
missing epoch, since every epoch is defined with a unique $\tau_i$ independent
of the others. As a result, cross-validation - a powerful
tool for model selection - is not possible.
% Removed this text to make the paper more concise at the request of the reviewer
%The key to understanding that point
%is that cross-validation essentially replicates the very process by which
%science operates; hypothesis-prediction-validation. We define some training
%subsample of the data, hypothesize a model including a regression of said model
%parameters, predict that model into the missing holdout data, and then ask
%whether it performs better than some null model (validation). No hypothesis can
%survive if it fails to make useful predictions.
Although cross-validation cannot be applied to model $\mathcal{T}$
directly, there is a way one can employ it. To do so, we work with the
marginalized transit times produced by model $\mathcal{T}$, rather than the
original photometry. This is less preferable for reasons
described earlier, but by doing so we can propose a simple sinusoidal model
against the derived transit times and use cross-validation to assess its
merit. In particular, we propose the following 5-parameter model for the
transit times:

\begin{align}
\tau(i) &= \tau_0 + i P + A_{\mathrm{TTV}} \sin(\nu_{\mathrm{TTV}} i + \phi_{\mathrm{TTV}}),
\label{eqn:taumodel}
\end{align}

where the TTV subscript terms control the sinusoidal feature of the model.
Note, that we applied our model to the transit times, not a list of TTVs. TTVs
are defined as deviations from a linear ephemeris, whose parameters are
themselves uncertain and indeed degenerate with the sinusoid, especially
for slow $\nu_{\mathrm{TTV}}$.

Of course, one could use this model on the photometry itself too (e.g.
see \citealt{ofir:2018}). However, cross-validation generally varies the choice
of training and hold-out sets, performing many realizations and then inspecting
the ensemble for the purposes of model comparison. Since our the photometric
fits take around a week to complete on $\sim$200 cores, it is not practical to
explore this approach in a reasonable time frame.

For our cross-validation, we defined a $20$\% hold-out set from the available
epochs. We then took the 80\% training set
and ran a weighted Lomb-Scargle periodogram \citep{lomb:1976,scargle:1982}
uniform in frequency. % from between the Nyquist rate to $1/(2B)$ where $B$
%is the baseline width of epochs available.
We selected the lowest-$\chi^2$ period and record the associated parameters. We also
performed a second fit with a simple linear ephemeris as the null model.
We then applied both models to the hold-out set and ask which one leads to
the best prediction in a $\chi^2$-sense\footnote{This implicitly means
we approximate the transit time posteriors as being Gaussian.}. We then
repeated the entire process, choosing another random group of
hold-out data, and continue $10^4$ times.

The cross-validation results are listed in
Table~\ref{tab:models}. A summary is that that none of the KOIs yield
cross-validation results where more than half of the sinusoidal predictions out
perform the linear ephemeris model, with the exception of KOI-303.01, which is
marginal at 54\%. However, KOI-303.01 was found earlier to not statistically
favour the existence of TTVs in a more general sense. This is because a) the
cross-validation results are marginal here and give almost even weight to the
competing hypotheses, and, b) the
earlier test treats the degrees of freedom as being equal to that of an
orbiting moon, but here the dimensionality is more restricted.

KOIs-1925.01 \& 2728.01 are worth commenting on since those
appeared to exhibit significant TTVs (see Section~\ref{sub:Q1}). As noted in the
previous subsection, the case for TTVs seems disparate between the first/second
halves of the data set for both objects and indeed the poor cross-validation
results make sense in this context. If there are stochastic TTVs (e.g. due
to stellar activity), a
deterministic model such as a sinusoid or exomoon will indeed fail to make
useful predictions, despite the fact that large and significant variations exist.

\subsection{Q3 - Are there moon-like timing variations?}
\label{sub:Q3}

The third and final question requires a model for the dynamical effect of
exomoons on the observations. It is not enough for a KOI to exhibit some kind
of TTVs, or a periodic TTV signal. This is because exomoons produce more subtle
and complex effects into the light curve than the approximate theory of
\citet{sartoretti:1999}, \citet{kipping:2009a} or \citet{kipping:2009b}. As
explicitly noted in \citet{thesis:2011}, expressions for the TTV (and TDV) waveform
caused by an exomoon, are approximate
and depend upon several assumptions. For example, the moon
and planet are assumed to experience no acceleration during the transit
duration, which requires that $P_S \gg T_{14}$. Given that the KOIs in question
have durations up to $\simeq$12 hours, %(as is true for KOIs-303.01, 1888.01 \&
%3320.01)
this implies moons less than few days orbital period would fail this
criteria. Further, exomoons induce other dynamical effects on the light curve
besides TTVs - such as TDV-Vs \citep{kipping:2009a}, TDV-TIPs \citep{kipping:2009b} and
ingress/egress asymmetry \citep{luna:2011}. Whilst TTVs are a sound place to
start an investigation, a detailed consideration of exomoon candidacy should
- in our opinion - consider the full details of the hypothesized model.

To address this then, we recommend a photodynamical analysis of the light
curve, which allows us to a) use full photometric time series, rather than
lossy derivative products; and b) fully model the subtle effects exomoons can
impart on the light curve.

Photodynamics models the light curve at each time step by evolving a $N$-body
system and calculating the fraction of stellar flux occulted to create a light
curve (e.g. see \citealt{barros:2015,almenara:2018,borkovits:2019}). In this
work, we use the \luna\ algorithm \citep{luna:2011} which is optimised for
exomoon fits and extends the \citet{mandel:2002} formalism.

The claim of \citet{fox:2020} is that these six KOIs exhibit transit timing
effects indicative of an exomoon. Transit timing effects are only sensitive
to the mass of an exomoon, not its radius; and thus, if the claim of
\citet{fox:2020} holds, then there should be some positive evidence for
a non-zero exomoon mass. The \citet{fox:2020} claim does not address
exomoon radius and so, even though that can be included in our photodynamical
model, we leave its inferred value aside for the time being and focus
on the photodynamically inferred exomoon mass.

Our moon model included the seven parameters from model $\mathcal{P}$
($P$, $\tau_0$, $p$, $b$, $\rho_{\star}$, $q_1$ \& $q_2$) as well as seven
additional satellite (``S'') parameters ($M_S/M_P$, $R_S/R_P$, $a_S/R_P$,
$P_S$, $\phi_S$, $\cos i_S$, $\Omega_S$). Note, that only six of these pertain
to the gravitational influence on the planet, and were thus counted as
penalized terms earlier in Section~\ref{sub:Q1}, since TTVs are not functionally
dependent on $R_S/R_P$. We adopted uniform priors for all terms
except for $P_S$, which has a log-uniform prior, and consider orbits out to
100 planetary radii. Models were regressed to the light curve using \multi,
as before.

If there are statistically significant transit timing effects (not just TTVs)
that were caused by an exomoon, then the exomoon mass in a photodynamical fit
would favour a non-zero value. In our fits, we were careful to not impose any
constraint on the exomoon density so that the posteriors can explore masses
tending to zero without penalization for unphysical satellite densities. The
planetary density, derived using the method of \citet{weighing:2010}, is
constrained to be 0.03\,g\,cm$^{-3} < \rho_P < 150$\,g\,cm$^{-3}$ to prevent
the code from exploring unphysical combinations of $P_S$ and $a_{SP}$.

\begin{figure}
\begin{center}
\includegraphics[width=8.4cm,angle=0,clip=true]{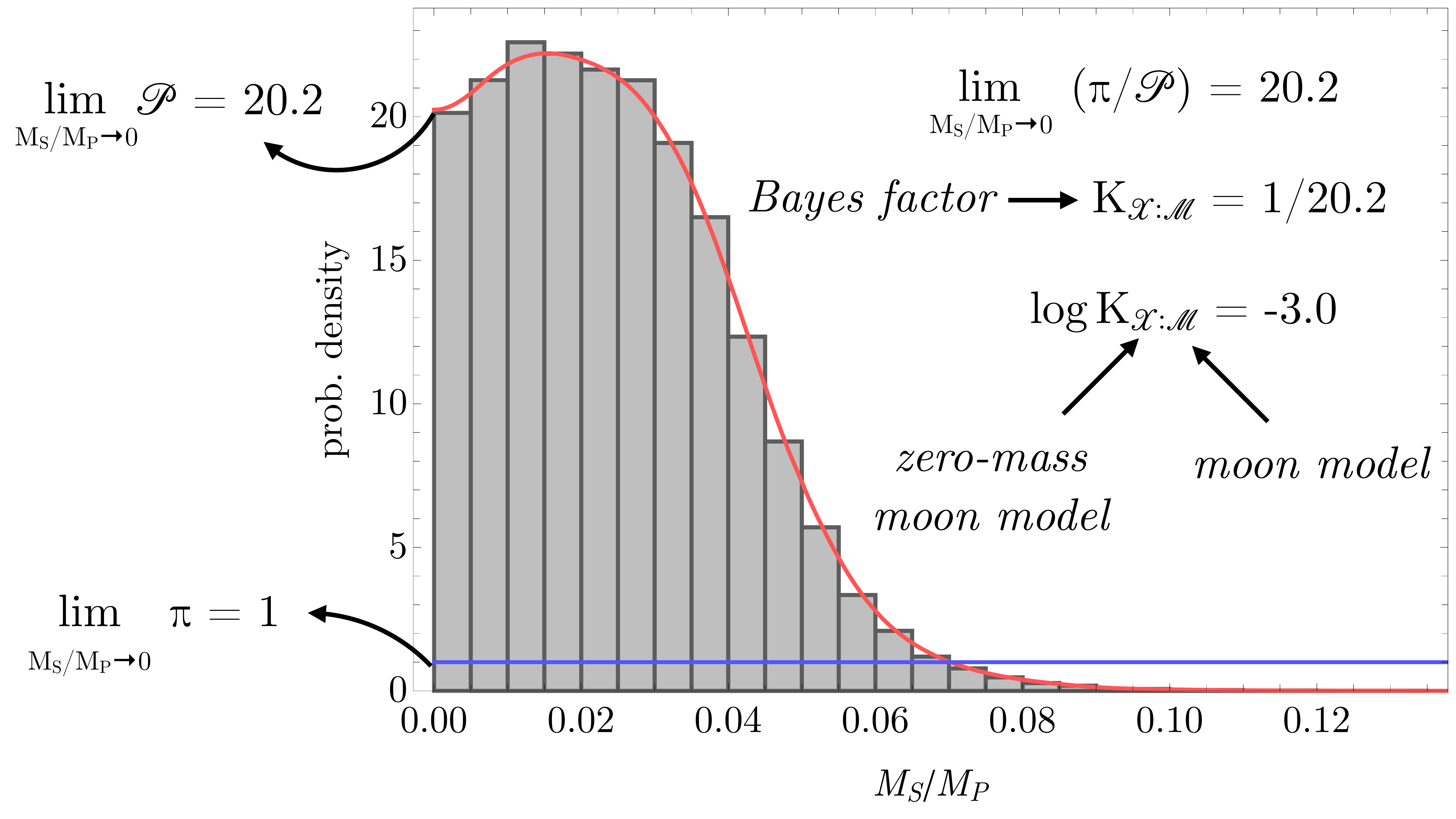}
\caption{
Example of how we calculated the SD ratio of the zero-mass moon model, here
for the case of KOI-1925.01. The histogram is calculated from the marginalized
posterior distribution, and this is generalized to a continuous function using
KDE (red line). We then evaluated the density in the limit of zero mass
and compare it to that of the prior. For example, here we find a 20:1 Bayes
factor in favor of a zero-mass moon model.
}
\label{fig:Msp}
\end{center}
\end{figure}

Mass is a positive definite quantity leading to traditional measures,
such as the median, to be become positively skewed, and thus posing a challenge
to straight-forwardly assessing its significance away from zero. To resolve
this, one might first consider using something like a \citet{lucy:1971} test,
but a more rigorous Bayesian approach is decribed in \citet{jontof:2015}
via the Savage-Dickey (SD) ratio \citet{dickey:1971}, and we follow that
approach here. We  evaluated the SD ratio by comparing the posterior density at
$M_S/M_P=0$ versus the prior (uniform) with an example illustrated in Figure~\ref{fig:Msp}.

The SD ratio allows for an estimate of the Bayes factor for
nested models.
Here, then, we compare the original full moon model, dubbed model
$\mathcal{M}$, against the same model but with no mass effects (dubbed
$\mathcal{X}$). Since the \citet{fox:2020} claim concerns transit timing effects
due to an exomoon, implaying an non-zero exomoon mass, then this
act directly evaluates the case for their claim in a Bayesian framework with
a self-consistent, photodynamical model.

Table~\ref{tab:models} shows the results of this exercise, where we find
a preference for zero-mass moon models for KOIs-268.01, 303.01, 1925.01 \&
3220.01 and very marginal preferences for a positive mass for KOI-1888.01
and KOI-2728.01.

\subsection{Other insights from the photodynamical fits}

Some other notable aspects of the results are briefly discussed. For KOIs
268.01, 303.01, 2728.01 \& 3220.01, the agreement between the light curve
stellar density and that from an isochrone analysis\footnote{This is achieved
by using the Gaia DR2 parallax, \kepler\ magnitude and \citet{mathur:2017}
DR25 stellar atmospheric properties of each KOI into \isochrones\
\citep{morton:2015}.} are within 2\,$\sigma$.
For KOI-1888.01, it's a little worse at 3\,$\sigma$. But for KOI-1925.01
the difference is pronounced, with the log of the ratio between them found
to be $\log(\rho_{\star,\mathrm{LC}}/\rho_{\star,\mathrm{isochrones}}) = 
2.2\pm0.3$, implying a minimum orbital eccentricity via the photoeccentric
effect \citep{dawson:2012} of $0.62\pm0.06$ - which would pose a significant
challenge for an exomoon due the truncation of the Hill sphere at periapse
\citep{domingos:2006}. We also verified this by taking the results from
model $\mathcal{T}$, evaluating a KDE of each segment's density ratio
posterior, taking the product of the two, numerically normalizing, and
then evaluating the median and standard deviation to give
$\log(\rho_{\star,\mathrm{LC}}/\rho_{\star,\mathrm{isochrones}}) = 
2.3\pm0.2$. On this basis, we assert with confidence that the densities
are in tension for KOI-1925.01 and the object likely maintains an
eccentricity in excess of $0.6$.

%Another useful quantity to track is $\rho_P$, the photodynamically inferred
%mean density of the planet. For KOI-268.01 \& KOI-1925.01, $\rho_P$ is
%unconstrained and spans the original prior range, and is thus not very
%revealing. For KOIs-303.01, 1888.01 \& 2728.01, the density pushes up against
%the lower prior limit, with 2\,$\sigma$ upper limits of
%$\rho_P < 2.3$\,g\,cm$^{-3}$, $<0.048$\,g\,cm$^{-3}$ and $<0.050$\,g\,cm$^{-3}$
%for each respectively. Since $\rho_P \propto a_{SP}^3 P_S^{-2}$, this behaviour
%indicates that fit ``wanted to'' make $a_{SP}/P_S$ as small as possible. For
%KOI-303.01, the density limit is quite sound, consistent with a mini-Neptune
%for example, but the other two densities are very low and raises some concern
%about that plausibility (although Kepler-51 provides a possible archetype
%example; \citealt{masuda:2014}).

We also note that our exomoon fits permit negative radius moons, which
translate to inverted transits and indeed some of our fits converge to
such unphysical solutions. In particular, KOI-268.01 \& 2728.01 both
strongly favour negative radius moons.

\begin{figure*}
\begin{center}
\includegraphics[width=17.0cm,angle=0,clip=true]{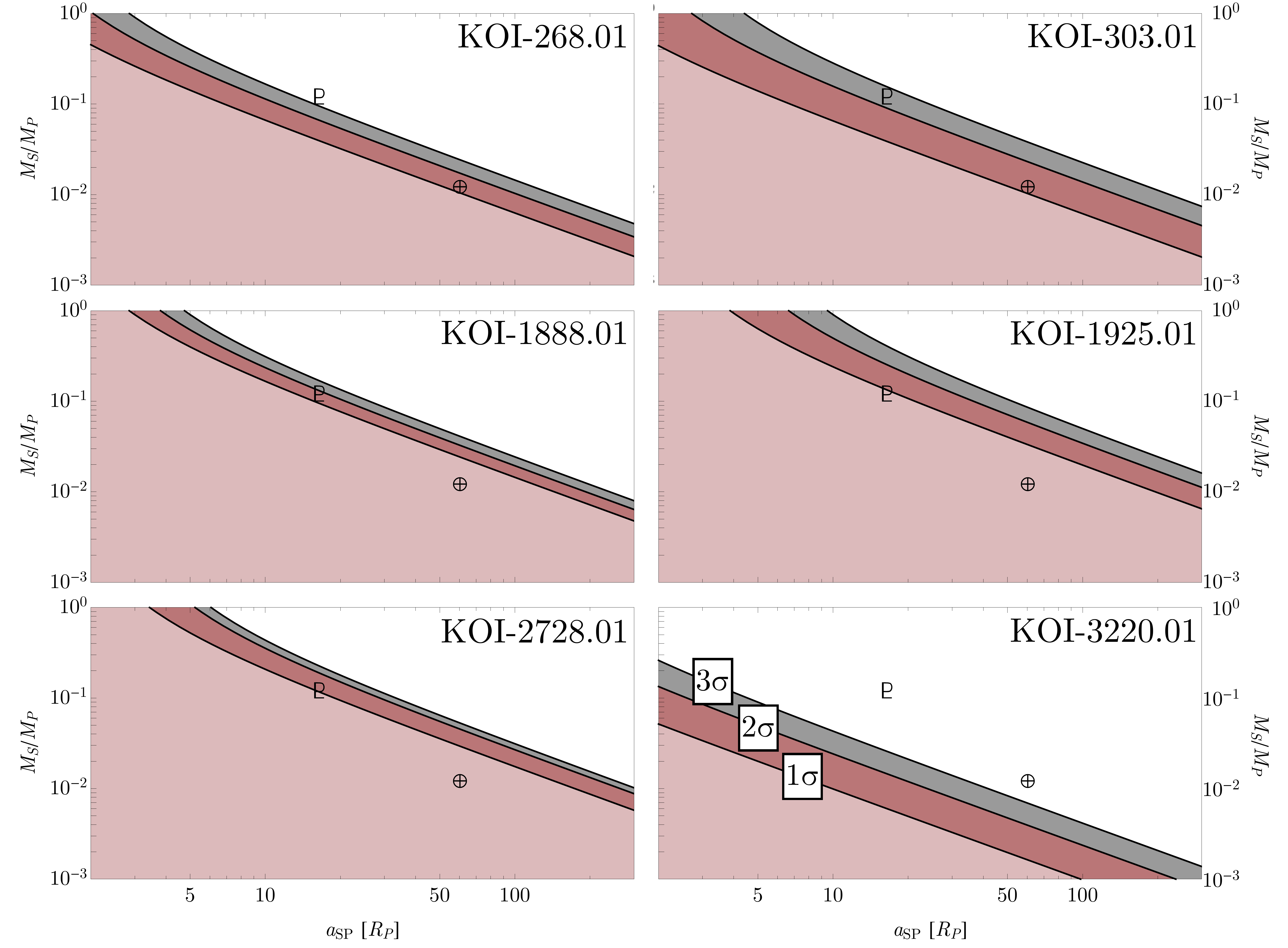}
\caption{
Mass limits to exomoons for KOIs-268.01, 303.01, 1888.01,
1925.01, 2728.01 \& 3320.01. Although we find no evidence for exomoon
candidates, the high signal to noise of these transits permits for
strong upper limits. We denote the position of Pluto-Charon and
the Earth-Moon on the diagram for context.
}
\label{fig:masslimits}
\end{center}
\end{figure*}

\subsection{Application to Kepler-1625b}

For completion, we decided to apply these tests to data used to claim an exomoon
candidate by \citet{teachey:2018}. The results are shown in Table~\ref{tab:models}.
Kepler-1625b passes Q1 and Q3 but we are able to evaluate Q2. The reason for this
is that with just four epochs, the act of regressing the five parameter model
given by Equation~(\ref{eqn:taumodel}) to the data leads to an over-determined
system. This is exacerbated if we drop an epoch for cross-validation purposes.
Nevertheless, we find that the results of the tests described here pose no
challenge to the candidacy of Kepler-1625b i.

%% file: modeltable.tex
\begin{table*}
\caption{
Statistical tests for evidence for exomoons, using gravitational effects only.
Answers to the three questions posed at the start of Section~\ref{sec:analysis}
are provided in columns 2, 3 \& 4, where the first number denotes the statistic
used to assess each question, and the mark in square brackets is a simple yes/no
summary of the posed question. The final column gives a mass ratio upper limit
derived.
We also show the same tests for Kepler-1625b \citep{teachey:2018}, although the
cross-validation test is not possible due to the limited number of samples, and
a mass upper limit is not provided since this case corresponds to a detection.
} % title of Table
\centering % used for centering table
\begin{tabular}{c c c c c} % centered columns (8 columns)
\hline\hline %inserts double horizontal lines
KOI &
$\mathrm{BIC}_{\mathcal{P}-\mathcal{T}}$ \textbf{[Q1]} &
\% of good TTV predictions \textbf{[Q2]} &
$\log K_{\mathcal{M}:\mathcal{X}}$ \textbf{[Q3]} &
$(M_S/M_P)$ at 60\,$R_P$ [2$\sigma$]\\ [0.5ex] % inserts table
%heading
\hline % inserts single horizontal line
268.01	& $-31.0$ [\text{\sffamily X}]	& 0\% [\text{\sffamily X}]	& $-3.7$ [\text{\sffamily X}]	& $<1.7$\% \\
303.01	& $-32.3$ [\text{\sffamily X}]	& 54\% [\checkmark]			& $-1.8$ [\text{\sffamily X}]	& $<2.3$\% \\
1888.01	& $-23.7$ [\text{\sffamily X}]	& 17\% [\text{\sffamily X}]	& $+0.6$ [\checkmark]			& $<3.3$\% \\
1925.01	& $+93.9$ [\checkmark]			& 41\% [\text{\sffamily X}]	& $-3.0$ [\text{\sffamily X}]	& $<5.9$\% \\
2728.01	& $+68.9$ [\checkmark]			& 28\% [\text{\sffamily X}]	& $+0.3$ [\checkmark] 			& $<4.6$\% \\
3220.01	& $-1.2$ [\text{\sffamily X}]	& 40\% [\text{\sffamily X}]	& $-6.9$ [\text{\sffamily X}]	& $<0.39$\% \\ 
\hline
K1625b & $+3.2$ & N/A & $+1.9$ & N/A \\ [1ex]
\hline\hline %inserts single line
\end{tabular}
\label{tab:models} % is used to refer this table in the text
\end{table*}

%% file: discussion.tex
In this work, we have conducted an independent examination of the claim of
\citet{fox:2020} that KOIs-268.01, 303.01, 1888.01, 1925.01, 2728.01 and 
3320.01 are ``exomoon candidates''. As the claim is based on transit timing
effects only, we have primarily framed our investigation in those same terms.

We structure our investigation in terms of three basic questions: 1) Are
there significant TTVs? 2) Is there a significant periodic TTV? 3) Is there
a statistically significant non-zero exomoon mass? It's worth noting that
the third criterion is a standard test used by the ``Hunt for
Exomoons with Kepler'' (HEK) project, namely criterion B2a \citep{HEK2,HEK5}.
Rather than rely on the catalog transit times of \citet{holczer:2016}, we
elected to infer our own times using method marginalized detrending of the
latest \kepler\ data products and incorporating short-cadence time series where
available.

The results of these three questions/tests are summarized in
Table~\ref{tab:models}. We find that KOIs-268.01 \& 3220.01 result in a ``no'' for
all three questions. KOIs-303.01, 1888.01 \& 1925.01 pass a single test each,
although a different one in each case. The analysis of this work thus concludes
that these five KOIs are not exomoon candidates.

Only KOI-2728.01 passes two of the three, failing the cross-validation test
when we ask if the periodic TTV has predictive capability. Specifically, when
we split the transit times into an 80:20 training:holdout set, we find that the
hypothesis of a periodic sinusoid defeats the predictions of the null
hypothesis (a linear ephemeris) in only 28\% of the draws. One explanation
would be that the TTVs are significant but are stochastic, perhaps caused by
stellar activity \citep{alonso:2008}, thus failing the periodic prediction
test. As an additional point of concern, KOI-2728.01 favours a negative-radius
moon when fit with a photodynamical exomoon model. On this basis, we do not
consider there to be a good case for KOI-2728.01 being an exomoon candidate.

It is important that we continue to search for exomoons, but they are unquestionably
very challenging objects to detect; not only at the hairy edge of \kepler's sensitivity,
but also plagued by a myriad of false-positives when considering a single
observable quantity, such as TTV. On the other hand their existence is equally
unquestionable, planets surely do have moons (!), but we caution that they demand very
high levels of care and statistical rigour. 

%% file: apptables.tex
\begin{table}
\caption{
Transit timing of KOI-268.01 derived in this work using model $\mathcal{T}$.
Transit times are quoted as $\mathrm{BJD}_{\mathrm{UTC}} - 2,455,000$. TTVs
are defined against the maximum a-posteriori ephemeris obtained from model
$\mathcal{P}$. Central values rethe median of the marginalized posterior
distribution and the uncertainties represents the $\pm$34.1\% range (TTVs
do not propagate the uncertainty of the ephemeris itself).
} % title of Table
\centering % used for centering table
\begin{tabular}{c c c} % centered columns (8 columns)
\hline\hline %inserts double horizontal lines
epoch &
$\tau_i$ &
TTV$_i$ [mins] \\ [0.5ex] % inserts table
%heading
\hline % inserts single horizontal line
$0$ & $8.9346_{-0.0020}^{+0.0020}$ & $1.5_{-2.9}^{+2.9}$ \\
$1$ & $119.3058_{-0.0023}^{+0.0026}$ & $-8.6_{-3.3}^{+3.8}$ \\
$2$ & $229.6835_{-0.0021}^{+0.0023}$ & $-9.4_{-3.0}^{+3.3}$ \\
$3$ & $340.0722_{-0.0020}^{+0.0021}$ & $5.8_{-2.9}^{+3.0}$ \\
$4$ & $450.4449_{-0.0011}^{+0.0011}$ & $-2.1_{-1.6}^{+1.6}$ \\
$6$ & $671.2044_{-0.0015}^{+0.0014}$ & $2.2_{-2.1}^{+2.0}$ \\
$7$ & $781.5830_{-0.0016}^{+0.0016}$ & $2.8_{-2.3}^{+2.4}$ \\
$8$ & $891.9602_{-0.0011}^{+0.0010}$ & $1.4_{-1.5}^{+1.5}$ \\
$9$ & $1002.3379_{-0.0014}^{+0.0013}$ & $0.7_{-2.0}^{+1.9}$ \\
$10$ & $1112.7123_{-0.0034}^{+0.0036}$ & $-4.8_{-4.9}^{+5.1}$ \\
$11$ & $1223.0886_{-0.0022}^{+0.0021}$ & $-7.5_{-3.2}^{+3.0}$ \\
$12$ & $1333.4798_{-0.0056}^{+0.0050}$ & $11.0_{-8.1}^{+7.2}$ \\ [1ex]
\hline\hline %inserts single line
\end{tabular}
\label{tab:koi268} % is used to refer this table in the text
\end{table}

\begin{table}
\caption{
Transit timing of KOI-303.01 derived in this work using model $\mathcal{T}$.
Transit times are quoted as $\mathrm{BJD}_{\mathrm{UTC}} - 2,455,000$. TTVs
are defined against the maximum a-posteriori ephemeris obtained from model
$\mathcal{P}$. Central values rethe median of the marginalized posterior
distribution and the uncertainties represents the $\pm$34.1\% range (TTVs
do not propagate the uncertainty of the ephemeris itself).
Horizontal line denotes the split between the two segments used.
} % title of Table
\centering % used for centering table
\begin{tabular}{c c c} % centered columns (8 columns)
\hline\hline %inserts double horizontal lines
epoch &
$\tau_i$ &
TTV$_i$ [mins] \\ [0.5ex] % inserts table
%heading
\hline % inserts single horizontal line
$0$ & $6.3664_{-0.0021}^{+0.0022}$ & $-6.2_{-3.1}^{+3.2}$ \\
$1$ & $67.2969_{-0.0022}^{+0.0021}$ & $-3.1_{-3.2}^{+3.1}$ \\
$2$ & $128.2240_{-0.0018}^{+0.0017}$ & $-4.9_{-2.6}^{+2.4}$ \\
$3$ & $189.1571_{-0.0020}^{+0.0021}$ & $2.1_{-2.9}^{+3.1}$ \\
$4$ & $250.0830_{-0.0025}^{+0.0025}$ & $-1.4_{-3.6}^{+3.5}$ \\
$5$ & $311.0143_{-0.0020}^{+0.0019}$ & $3.0_{-2.8}^{+2.8}$ \\
$7$ & $432.8734_{-0.0018}^{+0.0018}$ & $6.7_{-2.6}^{+2.7}$ \\
$10$ & $615.6579_{-0.0018}^{+0.0018}$ & $6.1_{-2.6}^{+2.6}$ \\
$11$ & $676.5786_{-0.0034}^{+0.0040}$ & $-4.9_{-4.9}^{+5.7}$ \\
$12$ & $737.5099_{-0.0024}^{+0.0023}$ & $-0.6_{-3.4}^{+3.3}$ \\
\hline
$13$ & $798.4419_{-0.0029}^{+0.0027}$ & $4.8_{-4.2}^{+3.8}$ \\
$14$ & $859.3700_{-0.0019}^{+0.0019}$ & $4.5_{-2.7}^{+2.8}$ \\
$15$ & $920.2974_{-0.0023}^{+0.0026}$ & $3.2_{-3.3}^{+3.7}$ \\
$16$ & $981.2224_{-0.0021}^{+0.0019}$ & $-1.4_{-3.0}^{+2.7}$ \\
$17$ & $1042.1503_{-0.0019}^{+0.0019}$ & $-2.0_{-2.7}^{+2.7}$ \\
$18$ & $1103.0801_{-0.0018}^{+0.0019}$ & $0.1_{-2.6}^{+2.7}$ \\
$19$ & $1164.0066_{-0.0017}^{+0.0017}$ & $-2.4_{-2.4}^{+2.4}$ \\
$20$ & $1224.9351_{-0.0018}^{+0.0018}$ & $-2.1_{-2.6}^{+2.6}$ \\
$21$ & $1285.8672_{-0.0021}^{+0.0019}$ & $3.3_{-3.1}^{+2.7}$ \\
$22$ & $1346.7855_{-0.0020}^{+0.0023}$ & $-11.0_{-2.9}^{+3.3}$ \\
$23$ & $1407.7215_{-0.0019}^{+0.0019}$ & $0.1_{-2.8}^{+2.8}$ \\ [1ex]
\hline\hline %inserts single line
\end{tabular}
\label{tab:koi303} % is used to refer this table in the text
\end{table}

\begin{table}
\caption{
Transit timing of KOI-1888.01 derived in this work using model $\mathcal{T}$.
Transit times are quoted as $\mathrm{BJD}_{\mathrm{UTC}} - 2,455,000$. TTVs
are defined against the maximum a-posteriori ephemeris obtained from model
$\mathcal{P}$. Central values rethe median of the marginalized posterior
distribution and the uncertainties represents the $\pm$34.1\% range (TTVs
do not propagate the uncertainty of the ephemeris itself).
} % title of Table
\centering % used for centering table
\begin{tabular}{c c c} % centered columns (8 columns)
\hline\hline %inserts double horizontal lines
epoch &
$\tau_i$ &
TTV$_i$ [mins] \\ [0.5ex] % inserts table
%heading
\hline % inserts single horizontal line
$0$ & $-32.8210_{-0.0037}^{+0.0035}$ & $-11.0_{-5.3}^{+5.0}$ \\
$1$ & $87.2100_{-0.0051}^{+0.0055}$ & $7.5_{-7.4}^{+7.9}$ \\
$2$ & $207.2265_{-0.0036}^{+0.0041}$ & $4.9_{-5.2}^{+5.9}$ \\
$3$ & $327.2412_{-0.0030}^{+0.0032}$ & $-0.2_{-4.3}^{+4.6}$ \\
$4$ & $447.2623_{-0.0029}^{+0.0029}$ & $4.0_{-4.2}^{+4.2}$ \\
$6$ & $687.3014_{-0.0032}^{+0.0033}$ & $7.8_{-4.6}^{+4.8}$ \\
$9$ & $1047.3508_{-0.0037}^{+0.0038}$ & $0.1_{-5.4}^{+5.5}$ \\
$10$ & $1167.3657_{-0.0030}^{+0.0031}$ & $-4.8_{-4.4}^{+4.5}$ \\
$11$ & $1287.3787_{-0.0035}^{+0.0035}$ & $-12.0_{-5.1}^{+5.0}$ \\
$12$ & $1407.4147_{-0.0035}^{+0.0035}$ & $13.0_{-5.1}^{+5.1}$ \\ [1ex]
\hline\hline %inserts single line
\end{tabular}
\label{tab:koi1888} % is used to refer this table in the text
\end{table}

\begin{table}
\caption{
Transit timing of KOI-1925.01 derived in this work using model $\mathcal{T}$.
Transit times are quoted as $\mathrm{BJD}_{\mathrm{UTC}} - 2,455,000$. TTVs
are defined against the maximum a-posteriori ephemeris obtained from model
$\mathcal{P}$. Central values rethe median of the marginalized posterior
distribution and the uncertainties represents the $\pm$34.1\% range (TTVs
do not propagate the uncertainty of the ephemeris itself).
Horizontal line denotes the split between the two segments used.
} % title of Table
\centering % used for centering table
\begin{tabular}{c c c} % centered columns (8 columns)
\hline\hline %inserts double horizontal lines
epoch &
$\tau_i$ &
TTV$_i$ [mins] \\ [0.5ex] % inserts table
%heading
\hline % inserts single horizontal line
$0$ & $12.0806_{-0.0045}^{+0.0026}$ & $-2.0_{-6.5}^{+3.8}$ \\
$1$ & $81.0356_{-0.0029}^{+0.0030}$ & $-7.0_{-4.2}^{+4.3}$ \\
$2$ & $150.0020_{-0.0027}^{+0.0026}$ & $4.5_{-3.8}^{+3.7}$ \\
$3$ & $218.9587_{-0.0052}^{+0.0054}$ & $2.0_{-7.6}^{+7.8}$ \\
$4$ & $287.9274_{-0.0026}^{+0.0026}$ & $17.0_{-3.7}^{+3.8}$ \\
$5$ & $356.8752_{-0.0026}^{+0.0031}$ & $1.3_{-3.7}^{+4.5}$ \\
$6$ & $425.8377_{-0.0022}^{+0.0025}$ & $7.2_{-3.1}^{+3.5}$ \\
$7$ & $494.7856_{-0.0028}^{+0.0023}$ & $-8.1_{-4.1}^{+3.3}$ \\
$9$ & $632.7067_{-0.0041}^{+0.0025}$ & $-2.0_{-5.9}^{+3.6}$ \\
$10$ & $701.6695_{-0.0042}^{+0.0037}$ & $4.2_{-6.1}^{+5.3}$ \\
\hline
$13$ & $908.5455_{-0.0023}^{+0.0015}$ & $5.1_{-3.3}^{+2.2}$ \\
$14$ & $977.49487_{-0.00077}^{+0.00095}$ & $-8.0_{-1.1}^{+1.4}$ \\
$15$ & $1046.4610_{-0.0012}^{+0.0012}$ & $3.0_{-1.8}^{+1.8}$ \\
$16$ & $1115.4145_{-0.0027}^{+0.0013}$ & $-4.1_{-3.8}^{+1.9}$ \\
$17$ & $1184.3791_{-0.0033}^{+0.0015}$ & $4.8_{-4.7}^{+2.1}$ \\
$18$ & $1253.3350_{-0.0015}^{+0.0014}$ & $1.2_{-2.1}^{+2.0}$ \\
$19$ & $1322.2931_{-0.0010}^{+0.0013}$ & $0.5_{-1.5}^{+1.9}$ \\ [1ex]
\hline\hline %inserts single line
\end{tabular}
\label{tab:koi1925} % is used to refer this table in the text
\end{table}

\begin{table}
\caption{
Transit timing of KOI-2728.01 derived in this work using model $\mathcal{T}$.
Transit times are quoted as $\mathrm{BJD}_{\mathrm{UTC}} - 2,455,000$. TTVs
are defined against the maximum a-posteriori ephemeris obtained from model
$\mathcal{P}$. Central values rethe median of the marginalized posterior
distribution and the uncertainties represents the $\pm$34.1\% range (TTVs
do not propagate the uncertainty of the ephemeris itself).
Horizontal line denotes the split between the two segments used.
} % title of Table
\centering % used for centering table
\begin{tabular}{c c c} % centered columns (8 columns)
\hline\hline %inserts double horizontal lines
epoch &
$\tau_i$ &
TTV$_i$ [mins] \\ [0.5ex] % inserts table
%heading
\hline % inserts single horizontal line
$2$ & $49.3312_{-0.0048}^{+0.0055}$ & $5.0_{-6.9}^{+7.9}$ \\
$4$ & $134.0463_{-0.0050}^{+0.0055}$ & $23.0_{-7.2}^{+7.9}$ \\
$5$ & $176.3782_{-0.0040}^{+0.0043}$ & $-4.5_{-5.7}^{+6.2}$ \\
$6$ & $218.7264_{-0.0047}^{+0.0042}$ & $-8.8_{-6.7}^{+6.1}$ \\
$7$ & $261.0789_{-0.0037}^{+0.0037}$ & $-6.9_{-5.3}^{+5.3}$ \\
$8$ & $303.4251_{-0.0051}^{+0.0054}$ & $-14.0_{-7.3}^{+7.8}$ \\
$9$ & $345.7901_{-0.0043}^{+0.0045}$ & $5.9_{-6.2}^{+6.4}$ \\
$12$ & $472.8512_{-0.0049}^{+0.0051}$ & $17.0_{-7.1}^{+7.4}$ \\
$13$ & $515.1917_{-0.0045}^{+0.0061}$ & $1.3_{-6.4}^{+8.7}$ \\
$15$ & $599.8986_{-0.0041}^{+0.0041}$ & $8.0_{-5.9}^{+5.9}$ \\
$16$ & $642.2450_{-0.0046}^{+0.0047}$ & $1.0_{-6.6}^{+6.8}$ \\
\hline
$17$ & $684.6047_{-0.0055}^{+0.0056}$ & $13.0_{-8.0}^{+8.1}$ \\
$18$ & $726.9712_{-0.0053}^{+0.0047}$ & $35.0_{-7.6}^{+6.7}$ \\
$21$ & $854.0108_{-0.0071}^{+0.0068}$ & $15.0_{-10.0}^{+9.7}$ \\
$23$ & $938.7014_{-0.0047}^{+0.0048}$ & $-1.6_{-6.8}^{+6.9}$ \\
$24$ & $981.0569_{-0.0039}^{+0.0038}$ & $4.5_{-5.6}^{+5.5}$ \\
$25$ & $1023.4057_{-0.0038}^{+0.0039}$ & $1.1_{-5.5}^{+5.6}$ \\
$26$ & $1065.7568_{-0.0048}^{+0.0041}$ & $1.0_{-6.9}^{+6.0}$ \\
$31$ & $1277.5145_{-0.0037}^{+0.0034}$ & $3.7_{-5.3}^{+4.9}$ \\
$33$ & $1362.2002_{-0.0038}^{+0.0034}$ & $-20.0_{-5.5}^{+4.9}$ \\
$34$ & $1404.5638_{-0.0044}^{+0.0042}$ & $-2.5_{-6.3}^{+6.1}$ \\ [1ex]
\hline\hline %inserts single line
\end{tabular}
\label{tab:koi2728} % is used to refer this table in the text
\end{table}

\begin{table}
\caption{
Transit timing of KOI-3220.01 derived in this work using model $\mathcal{T}$.
Transit times are quoted as $\mathrm{BJD}_{\mathrm{UTC}} - 2,455,000$. TTVs
are defined against the maximum a-posteriori ephemeris obtained from model
$\mathcal{P}$. Central values rethe median of the marginalized posterior
distribution and the uncertainties represents the $\pm$34.1\% range (TTVs
do not propagate the uncertainty of the ephemeris itself).
} % title of Table
\centering % used for centering table
\begin{tabular}{c c c} % centered columns (8 columns)
\hline\hline %inserts double horizontal lines
epoch &
$\tau_i$ &
TTV$_i$ [mins] \\ [0.5ex] % inserts table
%heading
\hline % inserts single horizontal line
$0$ & $-6.1363_{-0.0032}^{+0.0033}$ & $-1.7_{-4.6}^{+4.7}$ \\
$3$ & $238.1103_{-0.0024}^{+0.0022}$ & $-3.7_{-3.4}^{+3.2}$ \\
$4$ & $319.5344_{-0.0031}^{+0.0048}$ & $7.9_{-4.5}^{+6.9}$ \\
$5$ & $400.9478_{-0.0031}^{+0.0032}$ & $4.1_{-4.5}^{+4.6}$ \\
$6$ & $482.3613_{-0.0029}^{+0.0028}$ & $0.5_{-4.2}^{+4.1}$ \\
$8$ & $645.1897_{-0.0036}^{+0.0032}$ & $-4.7_{-5.1}^{+4.7}$ \\
$9$ & $726.6003_{-0.0025}^{+0.0024}$ & $-13.0_{-3.6}^{+3.5}$ \\
$10$ & $808.0274_{-0.0022}^{+0.0022}$ & $3.4_{-3.2}^{+3.1}$ \\
$11$ & $889.4434_{-0.0027}^{+0.0027}$ & $3.3_{-3.8}^{+4.0}$ \\
$14$ & $1133.6878_{-0.0024}^{+0.0025}$ & $-1.9_{-3.4}^{+3.6}$ \\
$15$ & $1215.0977_{-0.0024}^{+0.0025}$ & $-11.0_{-3.5}^{+3.6}$ \\
$16$ & $1296.5225_{-0.0023}^{+0.0024}$ & $2.1_{-3.3}^{+3.5}$ \\
$17$ & $1377.9448_{-0.0026}^{+0.0028}$ & $11.0_{-3.7}^{+4.0}$ \\ [1ex]
\hline\hline %inserts single line
\end{tabular}
\label{tab:koi3220} % is used to refer this table in the text
\end{table}